\documentstyle{article}
\newcommand{\be}{\begin{equation}}
\newcommand{\ee}{\end{equation}}
\newcommand{\bea}{\begin{eqnarray}}
\newcommand{\eea}{\end{eqnarray}}
\newcommand{\ba}{\begin{array}}
\newcommand{\ea}{\end{array}}

\def\bbox{{\,\lower0.9pt\vbox{\hrule \hbox{\vrule height 0.2 cm
\hskip 0.2 cm \vrule height 0.2 cm}\hrule}\,}}
\newcommand{\dsl}{\pa \kern-0.5em /}

\newcommand{\EQ}{\begin{equation}}
\newcommand{\EN}{\end{equation}}

\def\bbox{{\,\lower0.9pt\vbox{\hrule \hbox{\vrule height 0.2 cm
\hskip 0.2 cm \vrule height 0.2 cm}\hrule}\,}}

\newcommand{\pa}{\partial}

\def\today{\ifcase\month\or
  January\or February\or March\or April\or May\or June\or
  July\or August\or September\or October\or November\or December\fi
 \space\number\day, \number\year}

\input amssym.def
\input amssym.tex
\font\mybb=msbm10 at 10pt
\def\bb#1{\hbox{\mybb#1}}

\def\bR {\bb{R}}
\def\bE {\bb{E}}

\begin{document}


\begin{titlepage}
\vfill
\begin{flushright}
DAMTP-2002-133\\
hep-th/0211008\\
\end{flushright}


\vfill
\begin{center}
\baselineskip=16pt
{\Large\bf Surprises with Angular Momentum$^*$}
\vskip 0.3cm
{\large {\sl }}
\vskip 10.mm
{\bf ~Paul K. Townsend$^{\dagger}$ } \\
\vskip 1cm
{\small
$^\dagger$
DAMTP, University of Cambridge, \\
Centre for Mathematical Sciences,
Wilberforce Road, \\
Cambridge CB3 0WA, UK\\
}
\end{center}
\vfill
\par

\begin{center}
{\bf ABSTRACT}
\end{center}
\begin{quote}

The physics of angular momentum in even space dimensions can be surprisingly
counter-intuitive. Three such suprises, all associated with the properties
of supersymmetric rotating objects, are examined: (i) 5D black holes, (ii)
Dyonic instantons and (iii) Supertubes.

\vfill
 \hrule width 5.cm
\vskip 2.mm
{\small
\noindent $^*$ To appear in proceedings of TH-2002, The International
Conference on Theoretical Physics, Paris, UNESCO, July 2002.  \\
}

\end{quote}
\end{titlepage}

\setcounter{equation}{0}
\section{Prelude}

In his book {\it Surprises in Theoretical Physics}, Peierls examines various
occasions
on which his research led to a surprising conclusion \cite{RP} although, as
he says, 
{\sl there would have been no surprise if one had really understood the
problem
from the start}. Such  surprises are significant in that they expose flaws
in one's
physical intuition and thus serve to refine it. In the same spirit, this
article
recounts three surprising results of the author's co-investigations into the
properties of supersymmetric rotating objects. In each case one could
imagine a
similar surprise arising for non-supersymmetric objects, so supersymmetry
just
provides a convenient, and simplifying, context; the 
surprises are chiefly due to unexpected
features
of angular momentum, a fact that might itself be considered a meta-surprise
given the
central role of angular momentum in physics. However, intuition for angular
momentum
is usually acquired from a study of objects rotating in three-dimensional
space,
whereas the cases reviewed here involve rotation in either four space
dimensions or,
in the last case to be considered, two space dimensions. Although the
context of
each of the three surprises is quite different, there are a number of points
of
contact that make a comparative review seem worthwhile. I thank the
organisers of
TH-2002 for allowing me the opportunity to present such a review, and
I congratulate the Mayor of Paris for {\sl Paris Plage}.

\section{Supersymmetric Rotating Black Holes}

The first of our three surprises arose from a study of supersymmetric black
hole
solutions of 5D supergravity \cite{GMT}. Supersymmetric black holes are
special cases of stationary black holes. A stationary (asymptotically flat)
black hole
spacetime admits a Killing vector field (KVF) $k$ that is timelike near
spatial
infinity, and unique up to normalization. However, there may be interior
regions
outside the horizon, called `ergoregions', within which $k$ is spacelike; in
fact,
an event horizon with a non-zero angular velocity necesarily lies within an
ergoregion. Supersymmetric spacetimes cannot have ergoregions, however,
because
supersymmetry implies that $k$ can be expressed in terms of a Killing spinor
field,
and this expression allows $k$ to be timelike or null but not spacelike. It
follows
that {\it the event horizon of a supersymmetric black hole must be
non-rotating} \cite{GMT}.
This general observation, which applies in any spacetime dimension, should
be kept in
mind in what follows.

The angular momentum $J$ of any D-dimensional asymptotically-flat spacetime
can be
expressed as the surface integral
\be\label{jay1}
J = {1\over 16\pi G_D}\int_\infty dS_{\mu\nu} {\cal D}^\mu m^\nu
\ee
where $G_D$ is the D-dimensional Newton constant, ${\cal D}$ the standard
covariant
derivative, $m$ a suitably normalized spacelike KVF with
closed orbits and $dS$ the dual of the $(D-2)$-surface element at spatial
infinity,
which can be viewed as the boundary at infinity of a spacelike
$(D-1)$-surface
$\Sigma$. In the case of a black hole spacetime one may choose $\Sigma$ to
intersect
the event horizon on a $(D-2)$-surface $H$, in which case the expression
(\ref{jay1})
may be re-expressed in the form
\be
J= J_H + J_\Sigma
\ee
where $J_H$ is a surface integral over $H$ with the same integrand as in
(\ref{jay1})
and $J_\Sigma$ is a `bulk' integral over the region of $\Sigma$ outside the
horizon.

For solutions of 4D Einstein-Maxwell theory, and hence of pure $N=2$ 4D
supergravity, the integral $J_\Sigma$ vanishes and so $J=J_H$. In other
words, the
angular momentum is due to the black hole itself. This seems reasonable
given that
non-zero $J$ implies a rotating horizon by a theorem of Wald (which states
that a
stationary black hole with a non-rotating horizon is static \cite{RW}). But
a rotating
horizon is incompatible with supersymmetry so any 4D supersymmetric black hole
must be static.

The situation in 5D is more subtle. Firstly, the 5D Einstein-Maxwell theory
admits a
possible `$FFA$' Chern-Simons (CS) term, which is present in the pure 5D
supergravity theory with a particular coefficient \cite{Cremmer}. A
supersymmetric
black hole solution of this Einstein-Maxwell-CS theory must again have a
non-rotating
horizon but this no longer implies that $J=0$; there is no 5D analogue of
Wald's
theorem because the bulk Maxwell field may now carry angular momentum. In
fact, a
stationary supersymmetric black hole solution of 5D supergravity with
non-zero $J$ exists. It was first found in a slightly different context by
Breckenridge et al. \cite{BMPV} and is usually called the `BMPV' black 
hole. It was later shown to be a 1/2 supersymmetric solution of 
5D matter-coupled supergravity \cite{KRW,CS}. In the form
found in \cite{GMT}, as a solution of the pure minimal 5D
supergravity, the metric is
\be
\label{metric}
ds^2 = \left(1+ {\mu\over r^2}\right)^{-2} \left(dt + {j\sigma_3\over
2r^2}\right)^2 + 
\left(1+ {\mu\over r^2}\right)\left(dr^2 + r^2d\Omega_3^2\right)
\ee
where $d\Omega_3^2$ is the ($SU(2)_L\times SU(2)_R$)-invariant metric on
$S^3\cong
SU(2)$ and $\sigma_3$ is one of the three left-invariant forms on $SU(2)$
satisfying
$d\sigma_3 = \sigma_1\wedge\sigma_2$ and cyclic permutations. The parameters
$\mu$
and $j$ are related to the total mass $M$ and total angular momentum $J$ as
follows:
\be
\mu = {4MG_5\over 3\pi}\, ,\qquad j= -{2JG_5\over \pi}\, .
\ee 
The singularity at $r=0$ is just a coordinate singularity at a degenerate
non-rotating event horizon provided that
\be
j^2 < \mu^3\, .
\ee
Although the horizon has zero angular velocity, it {\sl is}
affected by the rotation; the horizon is a 3-sphere, topologically,
but geometrically it is a squashed 3-sphere, with a squashing parameter
proportional to $J$. As $j^2 \rightarrow \mu^3$ the squashed 3-sphere
degenerates, and for $j^2> \mu^3$ there are closed 
timelike curves through every point \cite{GMT}.
For this reason we restrict $j$ as above (and refer the interested reader to
\cite{GH,herd} for details of the `over-rotating' case).

Because the angular velocity of the horizon vanishes, one might expect to
find that
$J_H=0$ and hence that $J_\Sigma=J$. However, and this is the promised
surprise, a
calculation shows that \cite{GMT}
\be
\label{Jsig}
J_\Sigma = \left[1+ {1\over2}\left(1-{j^2\over\mu^3}\right)\right] J > J\, .
\ee
This implies not only that $J_H$ is non-zero but also that it is {\sl
negative}, as a
direct computation confirms! What this means is that a negative fraction of
the total
angular momentum is stored in the Maxwell field {\sl behind} the horizon.

Of course, given that there can be a contribution to the total angular
momentum of a
charged black hole from the Maxwell field {\it outside} the horizon there is
no good
reason to suppose that there is no similar bulk contribution from {\it
inside} the
horizon, and once this  has been appreciated it is not difficult to see why
the
fraction should be negative: given a positive bulk contribution
to the
angular momentum, one would expect frame dragging effects to cause the
horizon to
rotate unless these effects are counterbalanced by the frame dragging
effects due to a
negative contribution to the angular momentum in the fields behind the
horizon \cite{GMT}.
Because the horizon of a supersymmetric black hole cannot rotate,
$J_\Sigma>J$
should be expected. So why was it a surprise? The answer is presumably that
there is 
a clash with intuition derived from 
approaches to black hole physics such as
the
membrane paradigm \cite{TD,membrane} in which physical 
properties of the black hole
relevant to an exterior observer are expressed entirely in terms of the
horizon
and its exterior spacetime. Rotating 5D black holes appear to present an
interesting
challenge to this paradigm.

\section{Interlude: symmetries and angular momentum: I}

Before considering the next surprise it will be useful to consider the
effects, or
expected effects, of rotation on rotational symmetry. In four space
dimensions the
angular momentum 2-form $L$ has two skew eigenvalues $J \pm J'$, where $J$
and $J'$
are the quantum numbers associated to the rotation group $Spin(4) \cong
SU(2)_R \times
SU(2)_L$. One expects $spin(4)$ to be broken to $U(1)_R\times U(1)_L$ for
generic
$(J,J')$, but to $U(1)_R\times SU(2)_L$ when $J'=0$, in which case $L$ is
self-dual, and to $SU(2)_R\times U(1)_L$ when $J=0$, in which case
$L$ is anti-self-dual. The supersymmetric rotating 5D black hole has $J'=0$,
which is
why there is only a single rotation parameter $J$; the $U(1)_R\times
SU(2)_L$
symmetry is evident from the metric (\ref{metric}), and this is also
the isometry group of the squashed 3-sphere. 

The supersymmetry generators of minimal 5D supersymmetry transform as the
$({\bf 2},{\bf 1})\oplus ({\bf 1},{\bf 2})$ representation of $SU(2)_R\times
SU(2)_L$. Half the generators, call them $Q_L$, are singlets of 
$SU(2)_R$ and the other half, $Q_R$, are singlets of $SU(2)_L$. 
A half-supersymmetric configuration on which $Q_L$
acts trivially will preserve $SU(2)_L$ in which case the rotational
symmetry group must be unbroken or broken to $U(1)_R\times SU(2)_L$. 
In the case of the 5D black hole the former possibility applies 
when $J=0$ and the latter when $J\ne0$. More generally, 
we deduce that preservation of 1/2 supersymmetry
implies a self-dual or anti-self-dual angular momentum 2-form.

\section{Dyonic Instantons}

The instanton solution of Euclidean 4D Yang-Mills theory has an alternative
interpretation as a static soliton of the 5D Yang-Mills theory. As a
solution of 5D Super-Yang-Mills (SYM) theory these `instanton-solitons'
preserve 1/2 of the supersymmetry of the gauge theory vacuum. Let us
concentrate on
the minimal 5D SYM theory with gauge group $SU(2)$, for which the bosonic
field
content consists of the $SU(2)$ triplet of gauge potential 1-forms
$A^a$ ($a=1,2,3$) and a 
single scalar (Higgs) triplet $\phi^a$.
Supersymmetry does not permit a potential for the Higgs field so its
expectation
value is arbitrary. The vacua are thus parametrized by the constant 3-vector
\be
\langle \phi^a\rangle = v^a.
\ee
For vanishing Higgs field, and hence $v=0$, a Yang-Mills instanton is
a (marginally) stable
supersymmetric soliton solution of the 5D SYM theory with unbroken $SU(2)$.
A class
of multi-soliton solutions, with arbitrary instanton number $I$, is 
given by the 't Hooft ansatz
\be
A_i^a = \bar \eta_{ij}^a \partial_j \log H \qquad (i=1,2,3,4)
\ee
where $\bar \eta^a$ is the triplet of anti-self-dual complex structures on
$\bE^4$
and $H$ is a harmonic function on $\bE^4$ with point singularities such that
$H\rightarrow 1$ as $r\rightarrow \infty$, where $r$ is the radial distance
from the
origin of $\bE^4$. The simplest possibility,
\be
H= 1  + {\rho^2\over r^2}\, ,
\ee
yields the one-instanton ($I=1$) solution of `size' $\rho$.

When $v\ne0$ then $SU(2)$ is spontaneously broken to $U(1)$ and the
instanton-soliton
is destabilized; a simple scaling argument shows that the energy is reduced
if
$\rho$ is reduced, so the soliton will implode to a singular instanton
configuration
with $\rho=0$. However, the addition of an electric $U(1)$ charge can
stabilize the
soliton at some equilibrium radius, at which supersymmetry is again
partially
preserved \cite {LT}. Specifically, if we set
\be
A_0^a = \phi^a = v^a H^{-1}
\ee
then one again has a supersymmetric configuration. It is 1/2 supersymmetric
as a solution of the minimal SYM theory discussed here but 1/4
supersymmetric as a solution of the maximally-supersymmetric 5D SYM
theory, as is suggested by the formula 
\be
M = 4\pi^2|I| + |vq|
\ee  
for the mass of a dyonic instanton. 

The special case of
\be
H= 1 + {\rho\over r^2}
\ee
yields the one dyonic instanton ($I=1$) of size $\rho$. 
A computation of the $U(1)$
electric charge $q$ of this solution shows that $q\sim v/\rho^2$;
equivalently \cite{LT}
\be
\rho \sim \sqrt{q/v} \, .
\ee
The energy density takes the form
\be
{\cal E} = v^4 f(vr)
\ee
for some function $f$. As long as
\be
vq < 16\pi^2
\ee
the energy density is peaked about the origin in a region with diameter 
of order $\rho$. However, if
\be
vq > 16\pi^2
\ee
then the energy density takes its maximum on some 3-sphere centred on the
origin \cite{ETZ};
in the limit of large $vq$ the radius of this 3-sphere is of order $\rho$.

Note that the energy density of this $I=1$ solution 
is {\sl hyper-spherically-symmetric}, and the
3-sphere around
which the energy density is distributed at large $vq$ is a round 3-sphere,
not a
squashed one.  The rotational $Spin(4)\cong SU(2)_L\times SU(2)_R$ symmetry
{\sl is}
broken, to either $SU(2)_L$ or $SU(2)_R$, by the Yang-Mills-Higgs field
configuration
itself. However, these fields are also acted on by an `isospin' group
$SU(2)_I$, and
the diagonal subgroup $SU(2)_D$ of either $SU(2)_R\times SU(2)_I$ or
$SU(2)_L\times SU(2)_I$ survives. Thus the Yang-Mills-Higgs fields of the
dyonic
instanton preserve either $SU(2)_L\times SU(2)_D$ or $SU(2)_D\times
SU(2)_R$. This
is interpreted as the rotational invariance group of gauge-invariant
quantities such
as the energy density. Compare this state of affairs to that of the BPS
monopole of 4D N=2 SYM theory. In that case the rotation group
is $SU(2)$ and the group $SU(2)\times SU(2)_I$ is broken to 
the diagonal $SU(2)_D$ by the one-monopole solution, which is
therefore spherically symmetric. The one dyonic instanton solution is
hyper-spherically symmetric for essentially the same reason. 

The spherical symmetry of the BPS monopole indicates 
that it carries no angular momentum, because
angular momentum would break the spherical symmetry.  One might similarly
expect the
hyper-spherically symmetric dyonic instanton to carry no angular momentum
but, and
this is our second surprise, a computation yields \cite{ETZ}
\be
\label{ang2}
L_{ij} = -q \hat{\bf v}\cdot \bar\eta_{ij}
\ee
where $\hat{\bf v}$ is the unit 3-vector with components $v^a/v$. Thus, a
self-dual dyonic instanton has an anti-self-dual angular 
momentum 2-form proportional to $q$, and this is true even for the
hyper-spherically symmetric dyonic instanton solution of \cite{LT}.

The realization that the one dyonic instanton must carry angular
momentum, despite its hyper-spherical symmetry emerged from a
computation of its gravitational field \cite{ETZ}. Surprisingly, this turned out
to be stationary rather than static, and a subsequent
calculation of the angular momentum of the flat space dyonic instanton
yielded the above formula. Should this result not have been
anticipated from the start? Essentially, the
reason that the BPS monople has no angular momentum is that there is nothing
in the
monopole ansatz that could produce a non-zero result; the angular momentum
is zero because there is nothing else that it could be. Applied to the dyonic
instanton ansatz, the same argument shows only that $L \propto \hat{\bf v}\cdot
\bar\eta$, so the real question is why one should expect 
the constant of proportionality to vanish.
It vanishes when $q=0$ because the instanton is genuinely static, but when
$q\ne0$ we have electric fields and a configuration with electric fields is not
`genuinely' static (for a reason to be explained below) so the 
formula (\ref{ang2}) should not really have been a surprise.

Neverthess, I still find that the most common reaction to the statement {\it
spherical
symmetry in  four space dimensions does not imply vanishing angular
momentum} is
surprise. To mitigate the surprise I usually point out that 
circular symmetry in {\sl two} space
dimensions is obviously compatible with non-zero angular momentum. Pursuing
this point will lead to our third surprise.

\section{Interlude: symmetries and angular momentum: II}

A supersymmetric field configuration of a supersymmetric field theory is one
that is 
unchanged by the action of some linear combination $Q$ of supersymmetry
charges. This means that $Q^2$ acts trivially too, but the action of $Q^2$
on any
{\sl gauge-invariant} field is equivalent to the action of $H$, which
generates 
time translations. It follows that {\it a field configuration can be
supersymmetric
only if all gauge-invariant quantities are time-independent}; in particular,
this
means that the energy density must be time-independent\footnote{In the case
of
gravitational theories this argument needs modification because $Q$ is
defined only
as an integral at infinity, but the end conclusion is similar: {\it a
spacetime can
be supersymmetric only if it is stationary}.}. It is important to appreciate
that it
is possible for some {\sl gauge-dependent} field of a supersymmetric field
configuration to be time-dependent, and this is why there can exist
supersymmetric
field configurations with non-zero angular momentum. For example, a non-zero
electric
field ${\bf E}$ can have this effect because ${\bf E}=\dot {\bf A}$ in an
$A_0=0$
gauge, and in this gauge ${\bf A}$ is time-dependent if ${\bf E}$ is
non-zero. 

Although the energy density of a supersymmetric field configuration must be
time-independent, the possibility of a non-zero angular momentum indicates
that there
is motion nevertheless. Consider a circular planar loop of elastic string;
this may
rotate about the axis of the plane, and the associated angular momentum will
(if it
has a sufficiently large magnitude) support the string, against the force
exerted by its tension, at some equilibrium radius. A rotating configuration
of this
kind would not be incompatible with supersymmetry because the circular
symmetry ensures that the energy density of the string loop is
time-independent.
This illustrates the point that angular momentum is obviously compatible
with
circular symmetry in two space dimensions. 

It is even more obvious that
circular {\sl
asymmetry} is compatible with non-zero angular momentum but in this case it
might
seem unlikely that the energy density could be time-independent (as would be
required by supersymmetry); any rotating `bump' on the string would clearly
imply a
time-dependent energy density.

\section{Supertubes}

The supertube, as originally considered \cite{MT}, is a kind of string
theory
realization of the spinning string loop supported by angular momentum. I say
`kind of'
because a relativistic string that is described by a Nambu-Goto action
cannot support
momentum along the string and hence cannot rotate if it is circularly
symmetric.
However, IIA superstring theory has membrane solitons that appear as
D2-branes, and a
cylindrical D2-brane can be supported against collapse by angular momentum
in a
plane orthogonal to the axis of the cylinder. This is possible because the
D2-brane
action is not of Nambu-Goto type but rather of Dirac-Born-Infeld type and
the angular
momentum can be generated by the Born-Infeld (BI) electric and magnetic fields.
Specifically, the Lagrangian density is
\be
{\cal L} = -\sqrt{-\det(g+F)}
\ee
where $g$ is the induced metric on the 3D worldvolume and $F$
is the
worldvolume BI field strength 2-form. In principle we have a
membrane
in $\bE^9$ (since the spacetime is 10-dimensional) but we may choose to
consider a
membrane in $\bE^3\subset \bE^9$. A membrane of cylindrical topology can be
parameterized by worldspace coordinates $(z,\sigma) \in \bR\times S^1$. In a
physical gauge adapted to this topology, the geometry of a static membrane is
determined
by a single function $R(z,\sigma)$ which gives the radial position in the
plane
orthogonal to the axis of the cylinder as a function of position on the
brane. 
The BI magnetic field $B$ is a worldspace scalar. The BI electric field is a
worldspace 2-vector but, as we wish to generate an angular momentum in the
plane
orthogonal to the axis of the cylinder, we will choose this 2-vector to be
parallel 
to the axis of the cylinder. Thus, the BI 2-form is
\be
F= E(z,\sigma)\, dt\wedge dz + B(z,\sigma)\,  dz\wedge d\sigma\, .
\ee
The Lagrangian density now reduces to
\be
\label{lag0}
{\cal L} = -\sqrt{\left(R^2 + R_\sigma^2\right)\left(1-E^2\right) + B^2 +
R^2R_z^2}
\ee
where $R_\sigma = \partial_\sigma R$ and $R_z=\partial_z R$.

We have assumed that the D2-brane has cylindrical topology. If we further
assume that
it has cylindrical geometry then we must set $R_\sigma=0$ and $R_z=0$; the
radial
function $R$ thus becomes a real variable. The BI fields $E$ and $B$
similarly reduce
to real variables if we assume cylindrical symmetry, and the Lagrangian
density
becomes a function of three variables
\be
\label{lag}
{\cal L}(E,B,R) = -\sqrt{R^2(1-E^2) + B^2}\, .
\ee
Introducing the electric displacement
\be
{\cal D} \equiv {\partial {\cal L}\over \partial E}\, ,
\ee
the Hamiltonian density ${\cal H}= {\cal D}E-{\cal L}$ is
\be
{\cal H}(D,B,R) = R^{-1}\sqrt{\left(D^2 + R^2\right)\left(B^2 +
R^2\right)}\, .
\ee
This is equivalent to
\be
{\cal H}^2 = \left(D \pm B\right)^2 + \left({BD\over R} \mp
1\right)^2\, .
\ee
{}From this formula we see that the energy is minimised for given $B$ and $D$
when 
\be
R= |BD|.
\ee
This is therefore the equilibrium value of the cylinder radius.
The equilibrium
energy is 
\be
\label{enmin}
{\cal H}_{min} = |D| + |B|\, .
\ee
This energy formula is typical of 1/4 supersymmetric configurations, and a
calculation confirms that the D2-brane configuration just describes
preserves 1/4
supersymmetry, hence the name `supertube'.

As we go round the circle parametrized by $\sigma$, the tangent planes to
the tube
at a point with coordinates $(z,\sigma)$ are rotated by an angle in the
plane
orthogonal to the axis of the cylinder. Under normal circumstances a
configuration of
this type would not preserve supersymmetry because the D2 constraint on the
supersymmetry parameter associated with one tangent plane would be
incompatible with
the constraint associated with any of the other tangent planes. As an
extreme
example consider two tangent planes at diametrically opposite points on the
circle;
if we declare the constraint associated with one to be the D2-constraint
then the
constraint associated with the other is the anti-D2-constraint. Thus, the
supertube
is effectively a supersymmetric configuration that includes both D2-branes
and
anti-D2-branes!  There are various related ways to understand how this is
possible.
Note that the relation between $E$ and $D$ is
\be
E= {D\over R}\sqrt{B^2+R^2\over D^2+R^2}
\ee
and that this yields $E=1$ when $R=|BD|$. An electric field has the effect
of
reducing the D2-brane tension, and increasing $E$ to its `critical' value
$E=1$ would
reduce the tension to zero if the magnetic field were zero; this can be seen
from the
fact that ${\cal L}= -|B|$ for $E=1$. This explains why the supertube
has no
energy associated to the D2-brane tension; its energy comes entirely from
the
electric and magnetic fields, which can be interpreted as `dissolved'
strings and
D0-branes, respectively. The energy from the D2-brane tension has been
cancelled by
the binding energy released as the strings and D0-branes are dissolved by
the
D2-brane. Given that the D2-brane energy has been cancelled, it is perhaps
not so
surprising to discover that the D2-brane constraint is also absent, and
hence that
D2-branes can co-exist with anti-D2-branes without breaking supersymmetry. In
any
case, this is what happens and I refer to \cite{MT,NMT} for a much more
complete discussion of this point.

While the supersymmetry of the supertube might appear surprising, this
feature was
not discovered accidentally and, in any case, is not a surprise specifically
related
to angular momentum. Following the initial supertube paper \cite{MT}, a
matrix model
version of it was introduced by Bak and Lee \cite{BL}. A subsequent paper by
Bak and
Karch \cite{BK} found a more general solution of the matrix model describing
an
elliptical supertube, which included a plane parallel D2/anti-D2 pair as a
limiting
case. The fact that a circular cross-section could be deformed to an ellipse
was
certainly a surprise to me because it was hard to understand how any shape
other than
a circle could be consistent with both rotation and the time-independent
energy
profile required by supersymmetry (the limiting parallel brane/anti-brane
case, in
which angular momentum is replaced by linear momentum, seemed much less
problematical). And what was so special about an ellipse? Was this some
artefact of the matrix model approach? The possibility of a non-circular
cross-section had been considered, and rejected, in \cite{MT}, but what was
actually
shown there is that the cross-section must be circular if $R_z\ne0$, whereas
the
supertube has $R_z=0$. Let us return to (\ref{lag0}) and set $R_z=0$ but
keep the
variables $(E,B,R)$ as functions of $\sigma$. The same steps as before now
lead to
\be
{\cal H}^2 = \left(D \pm B\right)^2 + \left({BD\over \sqrt{R^2 + R_\sigma^2}
}\mp
1 \right)^2
\ee
As $D$ and $B$ are now functions of $\sigma$ we should minimise ${\cal
H}$ for fixed {\sl average} electric displacement $\bar D$ and {\sl average}
magnetic field $\bar B$, these quantities being proportional to the 
IIA string charge and D0-brane charge per unit length,
respectively. The result of this minimization procedure is that
suggested by the above formula; the energy 
is minimised when
\be
\sqrt{R^2 + R_\sigma^2} = |BD|\, ,
\ee
with {\sl no restriction} on $R(\sigma)$, and the energy at this minimum
is $|\bar D| + |\bar B|$. This implies preservation of 1/4
supersymmetry, as a direct computation confirms \cite{NMT}.
Note that the cross-sectional curve described by $R(\sigma)$ 
need not even be {\sl closed}; the net D2-brane
charge is non-zero for an open curve, which means 
that the supersymmetry algebra  
will include a D2-brane term in addition to the IIA string and 
D0 terms that are present for a closed curve. While this may be
another surprise it is not one related to angular momentum, so I refer 
to \cite{NMT} for details of its resolution. 

One way to understand how an arbitrary tubular cross-section can be
compatible with
supersymmetry is to note that the supertube is TST-dual to a wave of
arbitrary
profile on a IIA string \cite{NMT2} because it has been appreciated for a
long time
that any uni-directional wave on a string preserves supersymmetry
\cite{DGHW}.
However, this explanation provides little in the way of intuition that might
help us
understand the original problem: {\sl how can a rotating tubular D2-brane
with
a non-circular cross section have a time-independent profile}? There is a
simple explanation, which can be found in \cite{NMT}. I shall explain
it here in terms of a closely related undergraduate mechanics problem that I
heard about from Brandon Carter, whose earlier paper with Martin \cite{CM} 
describes a similar phenomenon in the context of superconducting 
cosmic strings. 

Consider an elastic hosepipe through which 
superfluid flows at variable velocity $v(s)$, where $s$
parametrizes the curve ${\cal C}$ described by the hosepipe.
If ${\cal C}$ is not straight then the motion of the fluid will produce  a
centrifugal force proportional to $v^2(s)/r(s)$ where
$r(s)$ is the radius of curvature of ${\cal C}$ at position $s$.
If we are free to choose the function $v(s)$ then we may choose it such that
the
centrifugal force produced by the fluid motion exactly balances, for all
$s$, the
centripetal force due to the hosepipe tension. Given ${\cal C}$, and a
parametrization of it, the function $v(s)$ will be determined by this local
force
balance condition. We may choose ${\cal C}$ to be closed, in which case the
hosepipe
forms a closed loop {\sl of arbitrary shape} that is prevented from collapse
by the
angular momentum generated by the circulating superfluid. A crucial feature
of this
hosepipe loop is that its shape does {\sl not} rotate, so the energy density
profile
is time-independent, despite the non-zero angular momentum. This shows that
time-independence does not imply circular symmetry.

\section{Epilogue}

I have discussed three surprises involving angular momentum in even space
dimension arising from various research projects on rotating supersymmetric
objects
undertaken over the past few years with Eduardo Eyras, Jerome Gauntlett,
David
Mateos, Robert Myers, Selena Ng and Marija Zamaklar. Although each surprise
arose in
its own distinct context, this article was motivated by the idea that there
is
something to be learnt by considering them together. Some of the connections
between
the three surprises have been discussed above, but there are others. For
example, the
supertube of circular cross-section is special in that it maximizes the
angular
momentum for given energy. This angular momentum upper bound arises in the
context of
the supergravity solution sourced by a supertube as a condition for the
absence of
global causality violations due to closed time-like curves \cite{EMT},
which is
precisely the origin of the upper bound on the angular momentum of a
supersymmetric
rotating 5D black hole. Also, the fact that the shape of a supertube does
not rotate
despite the non-zero angular momentum is reminiscent of the fact that the
horizon of
the `rotating' 5D black hole does not rotate despite its non-zero angular
momentum; of
course, the underying reason here is supersymmetry, but one wonders
whether the mechanisms might not also be related.

There are also several further connections between dyonic instantons and
supertubes. 
The dyonic instanton of 5D gauge theory has a 3D sigma-model analogue known
as a
Q-lump \cite{QL}; this is a charged sigma-model lump that expands to a loop
for large
charge in the same way that the dyonic instanton expands to a 3-sphere
\cite{AT}. 
In fact, it was this analogy that led to the realization that there should
exist a
tubular supersymmetric D2-brane supported by angular momentum; as explained
in
\cite{MT}, the supertube can be viewed as an effective worldvolume
description of the
sigma-model Q-lump. Given this, it is natural to wonder whether the dyonic
instanton
has a similar effective realization as a tubular D(3+p)-brane with a
3-sphere cross section (the 5D SYM/Higgs theory has a D4-brane realization, 
and the dyonic instanton then acquires a string-theory 
interpretation \cite{MZ} but this is quite different from the 
effective brane description being suggested here).

Finally, there is the question, which I leave unanswered, whether the
special
features of angular momentum in two and four space dimensions extend to
higher even
dimensions. For example, is non-zero angular momentum compatible with
$SO(2n)$ symmetry in $2n$ space dimensions for $n>2$?

\newcommand{\NP}[1]{Nucl.\ Phys.\ {\bf #1}}
\newcommand{\AP}[1]{Ann.\ Phys.\ {\bf #1}}
\newcommand{\PL}[1]{Phys.\ Lett.\ {\bf #1}}
\newcommand{\CQG}[1]{Class. Quant. Gravity {\bf #1}}
\newcommand{\CMP}[1]{Comm.\ Math.\ Phys.\ {\bf #1}}
\newcommand{\PR}[1]{Phys.\ Rev.\ {\bf #1}}
\newcommand{\PRL}[1]{Phys.\ Rev.\ Lett.\ {\bf #1}}
\newcommand{\PRE}[1]{Phys.\ Rep.\ {\bf #1}}
\newcommand{\PTP}[1]{Prog.\ Theor.\ Phys.\ {\bf #1}}
\newcommand{\PTPS}[1]{Prog.\ Theor.\ Phys.\ Suppl.\ {\bf #1}}
\newcommand{\MPL}[1]{Mod.\ Phys.\ Lett.\ {\bf #1}}
\newcommand{\IJMP}[1]{Int.\ Jour.\ Mod.\ Phys.\ {\bf #1}}
\newcommand{\JHEP}[1]{J.\ High\ Energy\ Phys.\ {\bf #1}}
\newcommand{\JP}[1]{Jour.\ Phys.\ {\bf #1}}

\end{document}